# Automation of Prosthetic Upper Limbs

# for Transhumeral Amputees Using

# Switch-controlled Motors

Fahad Moazzam Dar
College of E&ME
National University of Science and Technology (NUST)
Islamabad, Pakistan
fahaddar@gmail.com

Daniyal Malik
College of E&ME
National University of Science and Technology (NUST)
Islamabad, Pakistan
malik_daniyal@hotmail.com

Hassan Shahzad
College of E&ME
National University of Science and Technology (NUST)
Islamabad, Pakistan
mysterioushassan@yahoo.com

Umer Asgher
*SMME
National University of Science and Technology (NUST)
Islamabad, Pakistan
umer_asgher2000@yahoo.com

Emmad Adil
College of E&ME
National University of Science and Technology (NUST)
Islamabad, Pakistan
emmad_libra87@yahoo.com

Anees Ali
College of E&ME
National University of Science and Technology (NUST)
Islamabad, Pakistan
anees.ali2012@yahoo.com

*ABSTRACT*— The issues of "research required in the field of bio-medical engineering" and "externally-powered prostheses" are attracting attention of regulatory bodies and the common people in various parts of the globe. Today, 90% of prostheses used are conventional body-powered cable-controlled ones which are very uncomfortable to the amputees as fairly large amount of forces and excursions have to be generated by the amputee. Additionally, its amount of rotation is limited. Alternatively, prosthetic limbs driven using electrical motors might deliver added functionality and improved control, accompanied by better cosmesis, however, it could be bulky and costly. Presently existing proposals usually require fewer bodily response and need additional upkeep than the cable operated prosthetic limbs. Due to the motives mentioned, proposal for mechanization of body-powered prostheses, with ease of maintenance and cost in mind, is presented in this paper. The prosthetic upper limb which is being automated is for Transhumeral type of amputees i.e., amputated from above elbow. The study consists of two main portions: one is lifting mechanism of the limb and the other is gripping mechanism for the hand using switch controls, which is the most cost effective and optimized solution, rather than using complex and expensive myoelectric control signals.

## I. Introduction

N artificial limb is a type of prosthesis that replaces a missing extremity, such as arms or legs. The type of artificial limb used is determined largely by the extent of an amputation or loss and location of the missing extremity [1].

Ideally, prosthesis must be comfortable to wear, easy to put on and remove, lightweight, durable, and cosmetically pleasing. Furthermore, prosthesis must function well mechanically and require only reasonable maintenance. Finally, prosthetic use largely depends on the motivation of the individual, as none of the above characteristics matter if the patient doesn't want to wear the prosthesis [2].

This paper is an amalgamation of extensive research and novel automation of upper extremity. Many amazing advancements in prosthetic technology have been made in recent years, with more currently underway.

In general, robots are either controlled through remote control or programmed to perform certain tasks at a given time. A prosthetic limb, however, cannot be preprogrammed. by what means can we identify at what time a person will start to march down the staircases, or need to have a glass of water? Handheld controls will require a lot of attention along with the continuous usage of single or multiple hands. Two methods are presently employed for controlling the prostheses. One method is to utilize supplementary movements of the body, for instance using the movement of shoulders, to give a jerk to cables which in turn controls the wrist, hand, or elbow. Such type of prosthetic limbs are known as "cable-operated" prosthetic limbs. Small electrical pulses are produced by the muscles when they expand or contract. The more sophisticated technique is to extract these pulses and use the pulses from rest of the muscles to operate prostheses using motors. Small electrical antennas known as electrodes extract the pulses from muscles. For instance, somebody who is amputated beyond the elbow can utilize his triceps or biceps muscles to operate his artificial hand. The prosthetic limb can be preset to "open the hand" if it takes pulses from biceps muscle and "close the hand" if it takes pulses from triceps muscle. Such prosthetic limbs are termed as "myoelectric prosthetic limbs". Engineers of myoelectric prosthetic limbs have a main issue to resolve - actions to operate are many, but there is not sufficient muscles available to operate them. See the case in point above: if the biceps and triceps muscles are utilized to close and open the hand, which can be used to drive the pulses to elbow to straighten and bend it? It would be even more complex if the hand wishes to revolve sideward so as to get anything - how would the operator communicate it to carry it out? Scientists have tested numerous techniques to avoid such problems. They had employed numerous kinds of switches, or trained operators to execute arrangements of muscle expansion or contractions in place of switches. Nevertheless, the deficiency of accessible operating pulses, intricacy of design and struggle to control the operating pulses deem it exceedingly challenging to take myoelectric prostheses technology ahead [3].

The key purpose is to provide ease of use to the user and make the prosthetic upper limb cost effective using switch-controlled mechanism. When the user is unable to trained for myoelectric controls or sites producing such signals are not accessible, this could become a better option to operate the prosthetic limb. Mechanical switches would be placed under different easily accessible parts of the body, such as under armpits, which would act as control signals for the motorized arm. This would save many people from pain and misery because it would require fairly less amount of effort on part of the patient.

## II. Background

### A. Recent Technologies

In the modern era, developments in prosthesis have been noteworthy. Cutting-edge plastics and new materials, for example graphite, have made prostheses tougher and light in weight, restricting the quantity of additional power essential to control the prosthesis. This is especially important for transfemoral amputees. Additional materials have allowed artificial limbs to look much more realistic, which is important to transradial and transhumeral amputees because they are more likely to have the artificial limb exposed. In addition to new materials, the use of electronics has become very common in artificial limbs. Myoelectric and switch-controlled limbs allow the amputees to directly control the artificial limb. CAD and CAM are often used to assist in the design and manufacture of artificial limbs.

### B. Techniques Used

The variety of prostheses ranges from mostly passive or cosmetic types on one end to primarily functional types on the other. The aim of many prosthetic limbs lies in between.

Passive prosthetic limbs are also known as Cosmetic prostheses. Those amputees, who think their outlook is more essential, use these types of prostheses.

Functional prostheses generally can be classified into two categories. Cable-operated prosthetic limbs [4], [5] are typically less costly and generally have lighter mass. They have greater physical response and are highly robust prosthetic limbs. Yet, a cable-operated prosthetic limb is usually not visually attractive as compared to a myoelectric prosthesis. Also, the amputee has to apply a larger force in order to operate the prosthesis. Externally-powered prosthetic limbs [6], [7] are artificial extremities driven by electronically controlled motors and they deliver greater functionality and stronger grasping power, accompanied by better outlook, yet again they could be hefty and costly. These prosthetic limbs are operated using motors and batteries controlled by the amputees. At present, existing prostheses usually have lower physical response and need added upkeep as compared to cable-operated prosthetic limbs. A control system is always needed for such type of prosthetic limbs.

There are basically two kinds of control systems generally employed which are switch and myoelectric control. Muscle expansions and contractions are used as a pulse to operate the myoelectric prosthetic limb. The other type of control system employed in externally-powered prosthetic limbs is switch-controlled. It employs tiny switches, instead of pulses generated from muscles, to control the electronic motors. A switch can be activated by the movement of a remnant digit or part of a bony prominence against the switch. Each type of prosthesis has its own pros and cons depending upon amputation level (See Figure 1), patient's comfort, cost, weight and maintenance.

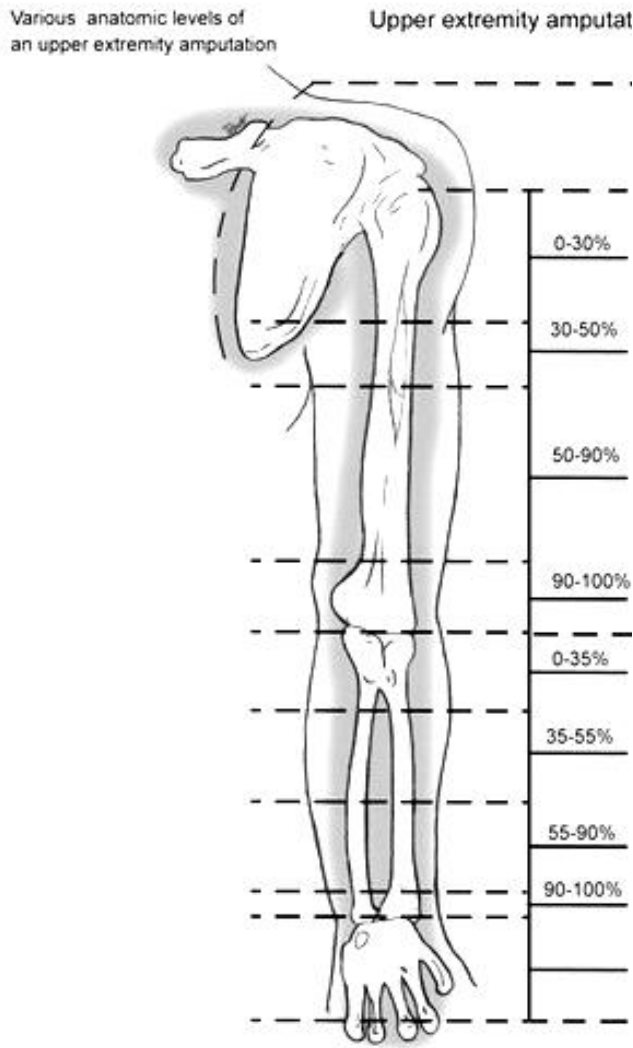

**Figure 1.** Amputation Levels [9]

## III.    System Overview

### A.    Description

The case under study is of Transhumeral amputation i.e., the patient is amputated from above elbow. For that purpose, we have divided the study into two portions i.e., working of elbow and grip.

#### 1.    Elbow Portion

In elbow part of the research a worm gear mechanism is used to lift the limb. For that purpose a very high torque and light weight motor will be used.

#### 2.    Grip Portion

In grip portion a worm gear mechanism is used with bidirectional motor control with further implementation of current feedbacks for limiting the constraints for motor.

### B.    Hardware Used

The system contains the following vital hardware components:

#### 1.    Mechanical Hardware

The mechanical hardware includes motors and switches mounted on the upper prosthetic extremity (See Figure 2 and Table 1). Design of the limb with placement and actuation of motors is thoroughly worked out. Actuation will be done using mechanical switches placed under both the armpits and behind the shoulder opposite to the arm which is amputated. One of our aims was to make the project cost effective. For that purpose, light weight, low cost and high torque motors for elbow lifting and gripping are selected.

Appropriate DC geared motor for lifting the limb is mounted on the limb for elbow movement with worm gear mechanism operated in upward and downward direction using two switches. As for the wrist movement, it is controlled using switches behind the shoulder activated by giving a sudden jerk to it and current feedback from the motor to judge if the object is firmly picked (See Figure 3). In addition to that, limit switches for the actuation of motors are prepared with adequate padding, so that it may not be too sensitive or too hard to be pressed.

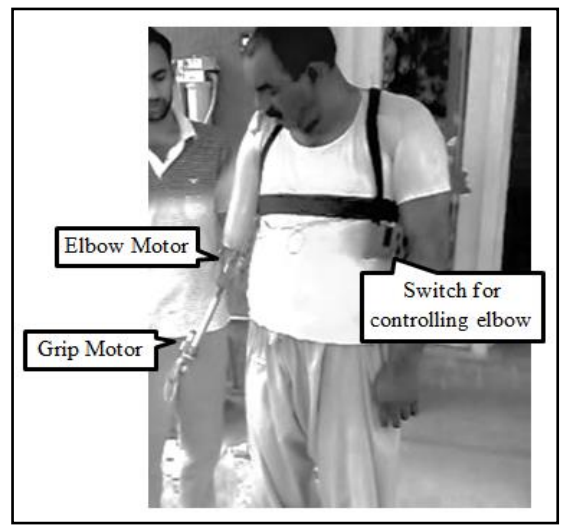

**Figure 2.** Complete System Fitted to the Amputee

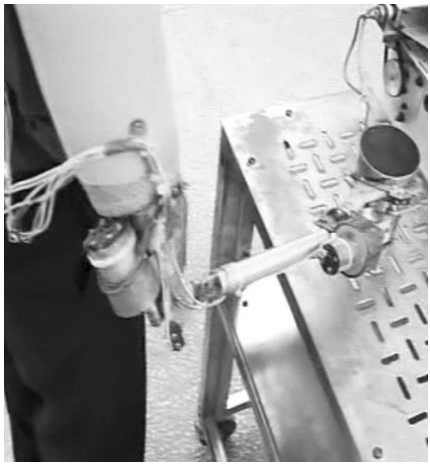

**Figure 3.** Lifting of an Object Using the Switch-controlled Prosthetic Limb

**Table 1.** List of Hardware Used

| Items | Quantity |
|---|---|
| Motor for lifting the limb | 1 |
| Motor for gripping | 1 |
| Switches for lifting the limb | 2 |
| Switch for gripping | 1 |

### 2. Electronics

The electronics include:

--A circuit for generic, h-bridge, voltage converter, ADC, and comparator.

--8051 microcontroller

--Battery

Electronics involved is simple and reliable and is tested thoroughly. Electronics of the project is complete with the block diagram (See Figure 4) and PCB layout (See Fig. 5) shown. Electronics consist of a single board with generic, 12V – 5V converter, h-bridge for driving the bidirectional motor, and power board on it. ADC and comparators are used for the current feedback from motor.

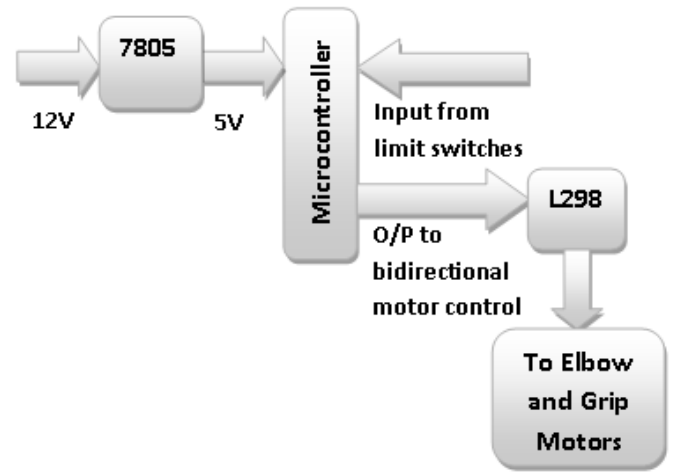

**Figure 4.** Block Diagram of Electronics

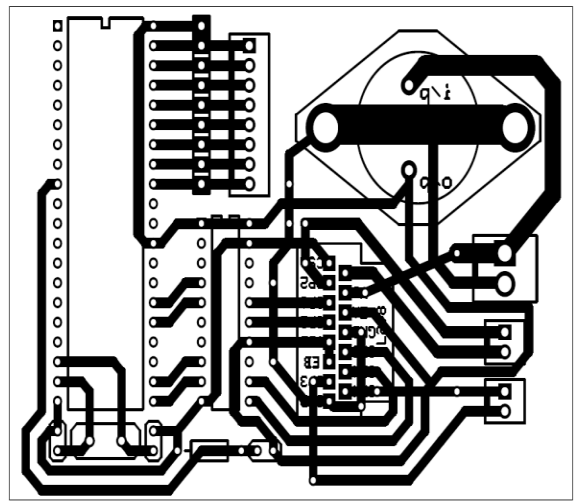

**Figure 5.** PCB Layout of Circuit

### C. Hardware Specification

### 1. Gripper Motor

The motor used for grip movement is 1271 bought from Mclennan Servo Supplies Ltd [10]. This DC motor is small in size and operates on a low speed which is perfectly suitable for an extensive variety of uses especially for driving a prosthetic limb (See Figure 6 and Figure 7). DC motor has iron core as its essential element which enables fluid action and an adjustable velocity in forward and reverse direction. The gear arrangement provides torque capable of 0.2 Nm (See Table 2).

The specifications indicate that this motor is quite appropriate for movement of the gripper. Also the unit is suitable for mounting on our prosthesis because of the small size. As our design requires small and low cost motors, this motor is ideal for that purpose.

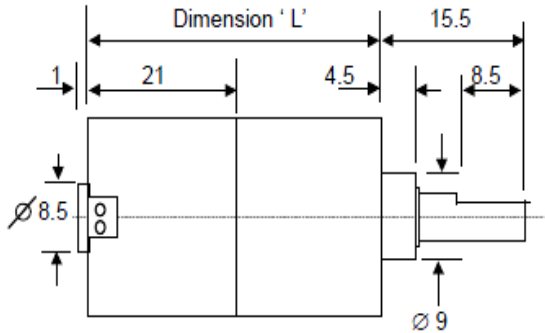

**Figure 6.** Motor's Front View. Note: All dimensions are in mm [10]

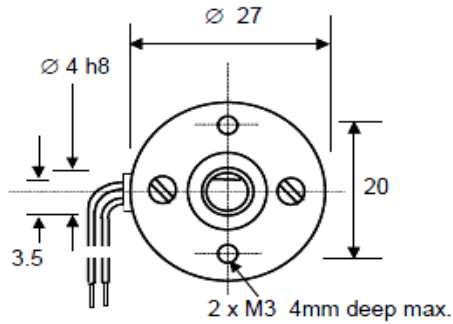

**Figure 7.** Motor's Top View. Note: All dimensions are in mm [10]

**Table 2.** Gripper Motor Specifications

| Specifications | |
|---|---|
| Length 'L' (mm) | 36 |
| Gear ratio | 10:1 |
| Nominal Voltage (Vdc) | 12 |
| No-Load speed (rpm) | 215 |
| Rated Speed (rpm) | 120 |
| Rated Torque (Nm) | 0.2 |
| Rated Current (mA) | 85 |
| Mass (grams) | 55 |

Note: Values are mentioned as quoted by the manufacturer. These are all approximate values [10]

### 2. Elbow Motor

The motor used for elbow movement is 80838.5 bought from Crouzet Ltd. The 80838.5 geared instrument DC motor is ideally suited to a wide range of applications requiring high torque, normal and low current. There is a ball bearing on drive shaft. Motor is light weight (See Table 3) with average size and relatively good gear ratio (See Fig. 8, Fig. 9 and Fig.

10). The motor is enclosed in a moulded aluminium case. It provides smooth operation and a bidirectional variable speed capability while the gearhead utilizes a metal spur gear which provides a very high torque for long life.

These dimensions and specifications show that the chosen motor is very suitable for elbow operation and could be mounted on the limb very easily. It is powerful enough to lift adequate weight as well as carrying the load of the limb and motor.

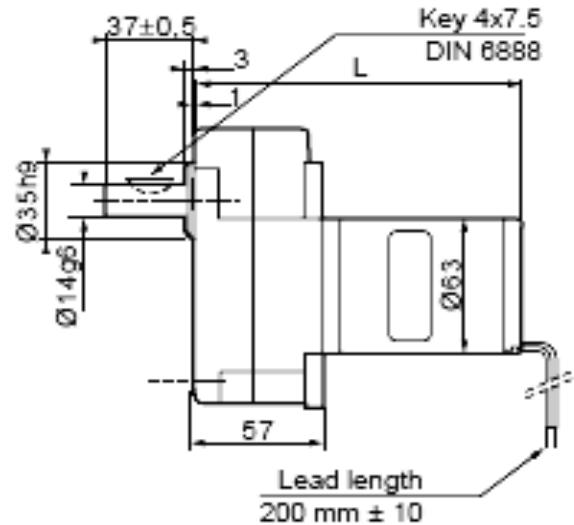

**Figure 8.** Elbow Motor's Front View. Note: All dimensions are in mm

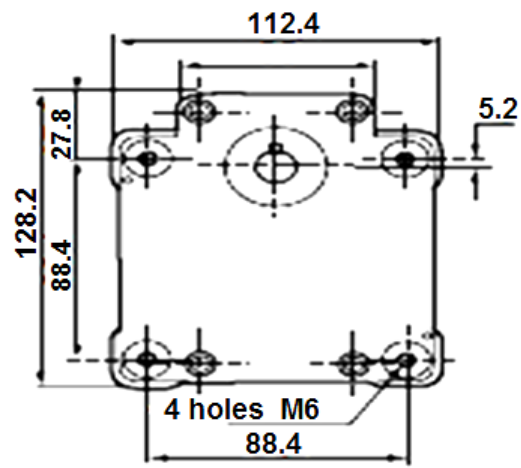

**Figure 9.** Elbow Motor's Top View. Note: All dimensions are in mm

**Table 3.** Elbow Motor Specifications

| Specifications | |
|---|---|
| Length 'L' (mm) | 85 |
| Gear ratio | 13:2 |
| Nominal Voltage (Vdc) | 24 |
| No-Load speed (rpm) | 135 |
| Rated Speed (rpm) | 80 |
| Rated Torque (Ncm) | 1.1 |
| Rated Current (mA) | 115 |
| Mass (grams) | 145 |

Note: Values are mentioned as quoted by the manufacturer. These are all approximate values.

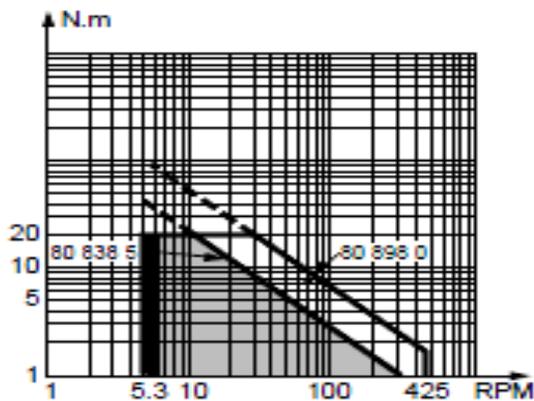

**Figure 10.** Nominal Speed and Torque Curves

### 3.  Body-powered Prosthetic Limb

The limb we are using to automate was bought from Otto Bock Ltd., Germany. The limb we got is a conventional body-powered prosthesis with a body harness used to operate the limb. This prosthetic limb comprises a socket (plastic-laminated and double-walled); a resistance wrist; a controlled open split hook; an adjustable pivot in elbow; a triceps cuff; a single control-cable system; and a harness frame (See figure 11).

The harness frame has two basic aims: control and suspension. The user can produce tightness in a cable used to control it via comparative physical actions. The tightness could be transmitted to the prosthesis, where it can generate the desired effect, by directing the cable via frame.

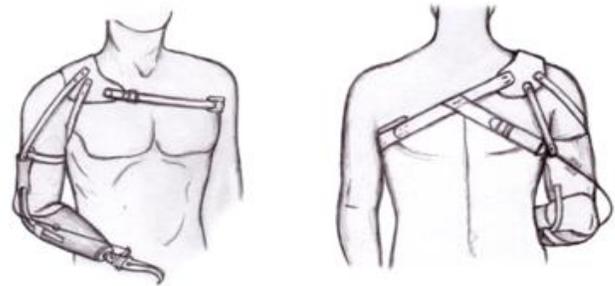

**Figure 11.** Body-powered Prosthetic Limb

#### D.  Microcontroller 8051

Two ports of microcontroller are used, one for input and the other one for output. Six pins of input and three pins of output port are used. Switches are used as inputs to the controller which in return controls the motors (See Table 4).

**Table 4.** Inputs and Outputs of Microcontroller

| SNo. | Inputs | Outputs |
|---|---|---|
| 1 | Elbow up limit switch | 2 pins for driving of elbow motor |
| 2 | Elbow down limit switch | 1 pin for driving of wrist motor |
| 3 | Grip open limit switch | |
| 4 | Maximum limit elbow up limit switch | |
| 5 | Maximum limit elbow down limit switch | |

Note: Limit switch for driving the elbow up will be placed under the right armpit and for driving the elbow down will be placed under left armpit. Limit switch for gripping will be placed below neck or behind the right shoulder which can be pressed by extending the arm. Interfacing of motors and switches with microcontroller is shown in Figure 4.

#### A.  Design Limitations

Efforts are made to make the product cost effective. Keeping this in mind the following compromises are to be made:

--Less degree of freedom as compared to myoelectric limb.

--Limb is controlled using mechanical switches instead of signals from body muscles.

--Limb is less cosmetically appealing.

--Extended initial testing required.

### IV.   Conclusion

The automation of limb is tested and implemented on the patient of Transhumeral type i.e., amputated from above the elbow joint. It proved to be a successful attempt towards a low cost and economically viable alternative. The effort showed that prosthesis can easily be operated using a switch-controlled mechanism which is fairly simple in design. Beneficiaries of this research include Armed Forces Institute of Rehabilitation and Medication (AFIRM) and their patients. Apart from that, this paper can benefit not only those who want to take up the field of research in bio-medical engineering but also the ones who are interested in providing easy to use and cost effective solutions in the field of prostheses.

References

[1]   Artificial limb. [Online], Retrieved on May 4, 20011.
      Available:
      http://en.wikipedia.org/wiki/Artificial_limb
[2]   A. Schultz, T. Kuiken, "New Prospects for Prosthetics." [Online], Retrieved on April 28, 20011.
      Available:
      http://scienceinsociety.northwestern.edu/content/articles/2009/kuiken/new-prospects-for-prosthetics
[3]   E. A. Biddiss and T. T. Chau, "Multivariate prediction of upper limb prosthesis acceptance or rejection (Report style)," Disabil Rehabil Assist Technol, Feb 10, 2008, pp. 1-12.
[4]   C. M. Fryer, G. E. Stark, J. W. Michael, "Body powered components(Book style with paper title and editor)," in *Atlas of Amputations and Limb Deficiencies,* 3rd ed., D. G. Smith, J. W. Michael, J. H. Bowker, Eds.  Rosemont, Ill: Am Acad Orthop Surg, 2004, pp. 117-30.
[5]   C. M. Fryer, J. W. Michael, "Harnessing and controls for body-powered devices(Book style with paper title and editor)," in. *Atlas of Amputations and Limb Deficiencies,* 3rd ed., D. G. Smith, J. W. Michael, J. H. Bowker, Eds.  Rosemont, Ill: Am Acad Orthop Surg, 2004, pp. 131-44.
[6]   C. W. Heckathorne, "Components for electric-powered systems(Book style with paper title and editor)," in. *Atlas of Amputations and Limb Deficiencies,* 3rd ed., D. G. Smith, J. W. Michael, J. H. Bowker, Eds. Rosemont, Ill: Am Acad Orthop Surg, 2004, pp. 145-72.
[7]   A. Muzumdar, "Powered Upper Limb Prostheses(Book style)," New York, NY: Springer, 2004.
[8]   A. Esquenazi, J. A. Leonard Jr, R. H. Meier, "Prosthetics, orthotics, and assistive devices(Book style with paper title and editor)," in *Arch Phys Med Rehabil,* May 1989, pp. 70-73.
[9]   B. M. Kelly, "Upper Limb Prosthetics eMedicine Physical Medicine and Rehabilitation" [Online], Retrieved on June 16, 20011.
      Available:
      http://emedicine.medscape.com/article/317234-overview
[10]  (Handbook style) Mclennan Servo Supplies Ltd. 1271 Series Datasheet [Online], Retrieved on June 16, 20011.
      Available:
      http://www.mclennanservosupplies.com/datasheets/european/geared/1271.pdf